\documentclass[twocolumn,10pt]{article}
\usepackage[utf8]{inputenc}
\usepackage[dvips]{graphicx}
\usepackage{amsthm}
\usepackage{amsmath}
\usepackage{amssymb}
\usepackage{natbib}

\newcommand{\ud}{\mathrm{d}}
\newcommand{\EdStime}{t_0^{\scriptscriptstyle{(EdS)}}}
\newcommand{\univage}{t_0^{\scriptscriptstyle{(\lambda)}}}

\title{Perturbatively constructed cosmological model with periodically distributed dust inhomogeneities}
\author{Szymon Sikora, Krzysztof Głód \\
\scriptsize{\textit{Astronomical Observatory, Jagiellonian University, Orla 171, 30-244 Kraków, Poland}} \\
\scriptsize{\textit{Copernicus Center for Interdisciplinary Studies, Szczepańska 1/5, 31-011 Kraków, Poland}} }

\begin{document}

\twocolumn[
\begin{@twocolumnfalse}
\maketitle
\begin{abstract}
We constructed a simple cosmological model which approximates the Einstein-de Sitter background with periodically distributed dust inhomogeneities. By taking the metric as a power series up to the third order in some perturbative parameter $\lambda$, we are able to achieve large values of the density contrast. With a metric explicitly given, many model properties can be calculated in a straightforward way which is interesting in the context of the current discussion concerning the averaging of the inhomogeneities and their backreaction in cosmology. Although the Einstein-de Sitter model can be thought as the model \emph{average}, the light propagation differs from that of the Einstein-de Sitter. The angular diameter distance-redshift relation is affected by the presence of inhomogeneities and depends on the observer's position. The model construction scheme enables some generalizations in the future, so the present work is a step towards more realistic cosmological model described by a relatively simple analytical metric. 
\end{abstract}
\vspace{1cm}
\end{@twocolumnfalse}
]

\small
\section{Introduction}

Inhomogeneous models of the Universe are an important tool of modern cosmology. They are widely used in the face of unsatisfying interpretation of astronomical observations within the framework of standard homogeneous Friedmann--Lema\^{i}tre cosmological model. The nature of dark energy which explains the accelerated expansion of the Universe in the concordance model is still unknown. Therefore, there arises a question if dark energy is not an artifact of inaccurately founded modeling which ignores possibly significant effect of inhomogeneities \cite{2011CQGra..28p4001E,2011CQGra..28p4009K,Clarkson:2011zq,2017IJMPD..2630011B}.

A recent discussion concerning a qualitative and quantitative influence of inhomogeneously distributed matter on the cosmological parameters of the Universe obtained with optical observations did not bring any coherent answer \cite{2014CQGra..31w4003G,2015PhRvL.114e1302A,2015CQGra..32u5021B,2015arXiv150606452G,2016CQGra..33l5027G,2017JCAP...03..062F,2017MNRAS.469L...1R,2017arXiv170609309A,2018A&A...610A..51R,2018PhRvD..97d3509E,2018ApJ...865L...4M}. This lack of consensus motivates studies of the light propagation in particular cases of inhomogeneous cosmological models. Among exact solutions to the Einstein field equations, models considered in this context are the Lema\^{i}tre--Tolman models \cite{2007JCAP...02..019E,2008JCAP...04..003G,2013CQGra..30i5011C} and their generalization, the Szekeres models \cite{2008PhRvD..78l3531I,2011PhRvD..84h9902I,2011JCAP...05..028N,2015PhRvD..91d3508K}. These models are studied with matter distributed not only in a~single halo cloud but also in various different ways, for example, in onion-like configurations \cite{2007JCAP...12..017B,2015PhRvD..92b3532K,2015PhRvD..92f9904K} or in layers of walls \cite{2012MNRAS.419.1937M,2012JCAP...02..036D}. Results of these studies can be compared with N-body relativistic simulations in a~weak field approximation \cite{2015PhRvD..91d3508K,2015PhRvD..92b3532K,2015PhRvD..92f9904K,2014PhRvD..89f3543A,2016PhRvD..93b3526A}. Furthermore, exact inhomogeneous models are used for building cosmological models in a~Swiss cheese arrangement which allows for light propagation studies in more realistic conditions \cite{2013PhRvD..87l3526F,2013JCAP...12..051L,2014PhRvD..90l3536P,2017PhRvD..95f3532K}. Another approach to observations in inhomogeneous cosmologies is based on models constructed as lattices of glued Schwarzschild regions \cite{2015PhRvD..92f3529L}, perturbatively solved point masses \cite{2013CQGra..30b5002B} or numerically simulated interacting black holes \cite{2017JCAP...03..014B}. There are also attempts to study light propagation in more versatile settings of post-Newtonian approach to gravitation \cite{2017JCAP...07..028S}. Recently, it becomes possible to analyze optical observations in cosmological models obtained with fully relativistic numerical simulations of space-time \cite{2016ApJ...833..247G}.

\begin{figure}[h]
	\centering
	\includegraphics[width=0.4\textwidth]{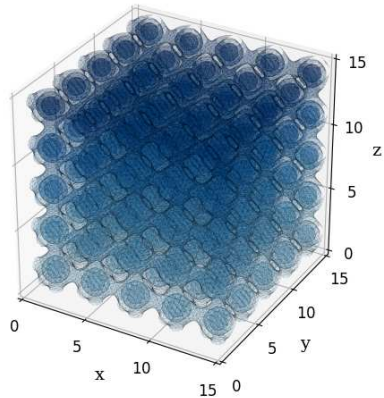}
	\caption{\label{fig:isodensity} \scriptsize{The model isodensity surfaces which form an infinite, periodic lattice.}}
\end{figure}

In our previous paper \cite{2017PhRvD..95f3517S} we constructed within the linear perturbation theory a simple model of the dust inhomogeneities on the Einstein-de Sitter (EdS) background. These inhomogeneities form an infinite, periodic, cubic lattice presented in Fig. \ref{fig:isodensity}. We have chosen such a density distribution because of the following reasons: \emph{(i)} For the volume much larger than the elementary cell the model becomes homogeneous and isotropic in common sense, so one can expect the Friedmann-Lema\^{i}tre-Robertson-Walker (FLRW) space-time as the average. \emph{(ii)} It satisfies the weak version of the Cosmological Principle, which means that there are no distinguished regions of the universe and each elementary cell looks the same. \emph{(iii)} The cosmological numerical simulations often adopt the cubic lattice with periodic boundary conditions, e. g. \cite{2016PhRvL.116y1302B,2016PhRvL.116y1301G}. Therefore, it is a need for analytical solutions with the same symmetries as tests of the numerical codes. 

The main goal of the present paper is to generalize our previous model beyond the first order of the linear perturbation theory so that a larger density contrast is allowed. The scheme presented here is open for further generalizations, so it is a single step towards the realistic inhomogeneous cosmological model. The paper is organized as follows. In Sec. \ref{sec:construction} the model details are presented. In Sec. \ref{sec:properties} we show some model properties and basic observables. Then, the conclusions follow. We presented in Sec. \ref{sec:construction} the approximate metric in the sense, that the resulting energy-momentum tensor is not exactly the dust one. It deviates from the dust solution in the second and the third order by some small, negligible terms. The advantage is that the simple, elementary functions appear in the metric elements. In Appendix, we present the strict dust solution up to second order. In that case, the metric tensor is much more complicated.

\section{The model construction}\label{sec:construction}
We assume the metric as a power series in some perturbative parameter $\lambda$:
\begin{equation}\label{ref:metric1}
g_{\mu\:\!\nu}=\sum\limits_{k=0}^{N}\,\lambda^k\,g^{(k)}_{\mu\:\!\nu}\,,
\end{equation}
where $g^{(0)}$ represents the EdS background metric. We adopt the cartesian coordinates\footnote{From now on, we will use the natural units $c=1$, $G=1$, and the convention in which Greek letters label the indices which cover the range $\{0,1,2,3\}$, while the Latin letters describe the space-like indices $\{1,2,3\}$} $\{t,x,y,z\}$ in which the background metric reads $g^{(0)}_{\mu\:\!\nu}=\mathrm{diag}(-1,a^2,a^2,a^2)$, where the scale factor is $a(t)=\mathcal{C}\,t^{2/3}$ and $\mathcal{C}$ is a constant. In each order $k\geq 1$, we introduce the comoving synchronous gauge, which means that $g^{(k)}_{\mu\:\!\nu}$ has the spatial part only: $g^{(k)}_{0\:\!0}=0$ and 
$g^{(k)}_{i\:\!0}=0$. This gauge condition guarantee that the vector $U^\mu=(1,0,0,0)$ is always tangent to the time-like geodesic. The dust particles follow the geodesic motion, so when the universe is filled with the dust, then the four-velocity $U^\mu$ represents the observer comoving with matter. 

For the metric (\ref{ref:metric1}) one can calculate the Einstein tensor $G_{\mu\:\!\nu}(\lambda)$. Let assume that the Einstein equations are satisfied. Then, the metric (\ref{ref:metric1}) is the model of space-time filled with the matter described by the energy-momentum tensor  $T_{\mu\:\!\nu}(\lambda)=G_{\mu\:\!\nu}(\lambda)/(8\pi)$. Our aim is to find such metric components $g^{(k)}_{\mu\:\!\nu}$, for which the tensor $T_{\mu\:\!\nu}(\lambda)$ approximates some physical energy-momentum tensor, in our case the dust one $T_{\mu\:\!\nu}^{(dust)}=\rho\,U_\mu\,U_\nu$. Let assume further that we may expand $G_{\mu\:\!\nu}(\lambda)$ in a Taylor series around $\lambda=0$: $G_{\mu\:\!\nu}(\lambda)=\sum\limits_{k=0}^{\infty}\,\lambda^k\,G^{(k)}_{\mu\:\!\nu}$. If there exists a similar expansion of the energy-momentum tensor: $T_{\mu\:\!\nu}(\lambda)=\sum\limits_{k=0}^{\infty}\,\lambda^k\,T^{(k)}_{\mu\:\!\nu}$, we can identify the elements $T^{(k)}_{\mu\:\!\nu}=G^{(k)}_{\mu\:\!\nu}/(8\pi)$. We will analyze the terms $T^{(k)}_{\mu\:\!\nu}$ order by order. If $T_{\mu\:\!\nu}(\lambda)\approx T_{\mu\:\!\nu}^{(dust)}$, then $T^0\,{}_0(\lambda)=-\rho$ and the other components of $T^\mu\,{}_\nu(\lambda)$ should be as close to zero as possible. For convenience, we will use a formal expansion of the density in the same form: $\rho=\sum\limits_{k=0}^{\infty}\,\lambda^k\,\rho^{(k)}$.

Now, in order to find the metric for which $T_{\mu\:\!\nu}(\lambda)$ approximates the dust with a distribution similar to that presented on the Fig. \ref{fig:isodensity}, we impose one symmetry condition. We demand that the metric $g_{\mu\:\!\nu}$ is invariant under every permutation of the spatial variables $\{x,y,z\}$. Then, we can consider a possible decomposition of the metric in a given order which takes the form:
\begin{multline}\label{Eqn:metric_form}
g^{(k)}_{i\:\!j}=
\begin{pmatrix}
A_x^{(k)} &0&0\\
0&A_y^{(k)}&0\\
0&0&A_z^{(k)}
\end{pmatrix}+
\begin{pmatrix}
B_{x\:\!y\:\!z}^{(k)} &0&0\\
0&B_{x\:\!y\:\!z}^{(k)}&0\\
0&0&B_{x\:\!y\:\!z}^{(k)}
\end{pmatrix}+\dots \\
\dots+
\begin{pmatrix}
C_{y\:\!z}^{(k)} &0&0\\
0&C_{x\:\!z}^{(k)}&0\\
0&0&C_{x\:\!y}^{(k)}
\end{pmatrix}+ 
\begin{pmatrix}
0&D_{z}^{(k)}&D_{y}^{(k)}\\
D_{z}^{(k)}&0&D_{x}^{(k)}\\
D_{y}^{(k)}&D_{x}^{(k)}&0
\end{pmatrix}+\dots \\
\dots+
\begin{pmatrix}
0&E_{x\:\!y\:\!z}^{(k)}&E_{x\:\!y\:\!z}^{(k)}\\
E_{x\:\!y\:\!z}^{(k)}&0&E_{x\:\!y\:\!z}^{(k)}\\
E_{x\:\!y\:\!z}^{(k)}&E_{x\:\!y\:\!z}^{(k)}&0
\end{pmatrix}+
\begin{pmatrix}
0&F_{x\:\!y}^{(k)}&F_{x\:\!z}^{(k)}\\
F_{x\:\!y}^{(k)}&0&F_{y\:\!z}^{(k)}\\
F_{x\:\!z}^{(k)}&F_{y\:\!z}^{(k)}&0
\end{pmatrix}\,,
\end{multline}
where we used the abbreviations $A_i^{(k)}\equiv A^{(k)}(t,x^i)$, $B_{i\:\!j\:\!l}^{(k)}\equiv B^{(k)}(t,x^i,x^j,x^l)$, $C_{i\:\!j}^{(k)}\equiv C^{(k)}(t,x^i,x^j)$, $D_{i}^{(k)}\equiv D^{(k)}(t,x^i)$, $E_{i\:\!j\:\!l}^{(k)}\equiv E^{(k)}(t,x^i,x^j,x^l)$, $F_{i\:\!j}^{(k)}\equiv F^{(k)}(t,x^i,x^j)$. In this setup, to specify the metric in a given order, one should propose six, symmetric in spatial coordinates functions $A^{(k)},\dots,F^{(k)}$.

\subsection{The first order}\label{sec:FirstOrder}
In the notation introduced above the metric from our previous paper \cite{2017PhRvD..95f3517S} can be recovered by putting the only nonzero function in the first order as:
\begin{equation}\label{Eqn:Metric1}
A^{(1)}(t,w)=-\,t^{-1}\,\EdStime\,\sin^2(\mathcal{B}\,w)\,,
\end{equation}
where $\EdStime=2/(3H_0)=9.32\,\mathrm{Gyr}$ is the age of the EdS universe for the Hubble constant $H_0=70\,\mathrm{km/s/Mpc}$ and $\mathcal{B}$ is the parameter related to the size of the overdensities. For such a choice one can show that $T^{(1)\:\!0}\,{}_0=-\rho^{(1)}$, where:
\begin{equation}
\rho^{(1)}=\frac{2\,t_0^{(EdS)}}{3\,t^3}\left(\sin^2(\mathcal{B}\,x)+\sin^2(\mathcal{B}\,y)+\sin^2(\mathcal{B}\,z) \right)\,,
\end{equation}
and other components of $T^{(1)\:\!\mu}\,{}_\nu$ are exactly zero. Since $\rho^{(0)}=(4/3)\,t^{-2}$, up to the first order $T_{\mu\:\!\nu}(\lambda)$ represent an exact dust solution, for which the periodically distributed inhomogeneities have the amplitude decreasing in time.

\subsection{The second order}
To estimate the contribution to the energy-momentum tensor from the higher orders we have to specify the values of the model parameters. We will use the Mpc as a unit of length. Then, in the natural units convention $c=1$, the age of the EdS universe is $\EdStime=2855.16\,\mathrm{Mpc}$. For the typical scaling $a(\EdStime)=1$, the value of the constant $\mathcal{C}$ is $4.96\times 10^{-3}$.
The elementary cell $\mathcal{D}$ is the region $x\in(0,\pi/\mathcal{B}),y\in(0,\pi/\mathcal{B}),z\in(0,\pi/\mathcal{B})$. For the simple choice $\mathcal{B}=1$, the size of the elementary cell should be around $3\,\mathrm{Mpc}$, which is a typical scale of the galaxy clusters. Further, it is convenient to measure the energy-momentum contribution in the units of the critical density $\rho_{cr}=(3H_0^2)/(8\pi)$. We introduce the definitions $\Omega^{(k)}:=\rho^{(k)}/\rho_{cr}$ and $\Omega^{(k)\:\!\mu}\,{}_\nu:=T^{(k)\:\!\mu}\,{}_\nu(\lambda)/\rho_{cr}$. Of course, the background density of the EdS model at $t_0^{(EdS)}$ is $\Omega^{(0)}=1.0$. Let's take the relatively high amplitude $\lambda=4/15\approx0.26$. Then, at the maximum of the overdensity $\vec{x}_{\mathcal{O}}=(\frac{\pi}{2},\frac{\pi}{2},\frac{\pi}{2})$ at the time $t_0^{(EdS)}$, we have $\Omega^{(1)}=0.4$. The observer located in the underdensity $\vec{x}_{\mathcal{U}}=(0,0,0)$ measures $\Omega^{(1)}=0.0$, so the density contrast is quite high.

For the metric of the form (\ref{Eqn:metric_form}) the energy-momentum tensor in each order $k\geq 1$ has the four types of components: $T^{(k)\:\!0}\,{}_0$, $T^{(k)\:\!i}\,{}_0$, $T^{(k)\:\!i}\,{}_j$ for $i=j$ and $T^{(k)\:\!i}\,{}_j$ for $i\neq j$. Each type characterizes by the same structural dependence on the metric functions. For the amplitude $\lambda$ given above, the maximum over the elementary cell $\mathcal{D}$ of the diagonal elements in the second order is $\max\limits_{x^\mu\in\mathcal{D}}|\Omega^{(2)\:\!i}\,{}_j|_{i=j}=0.013$. It is a matter of luck that one can easily get rid of these elements with the help of the $A^{(2)}$ and $F^{(2)}$ as the only nonzero metric functions in the second order, by putting:
\begin{multline}\label{Eqn:Metric2}
A^{(2)}(w)=t^{-2}\,(\EdStime)^2\,\left(\frac{\sin^4(\mathcal{B}\,w)}{4}-\frac{\sin^2(\mathcal{B}\,w)}{8}+\frac{1}{32} \right)\,,\\
F^{(2)}(w_1,w_2)=-\frac{t^{-8/3}\,(\EdStime)^2\,\mathcal{C}^2}{64\,\mathcal{B}^2}\,\sin(2\,\mathcal{B}\,w_1)\,\sin(2\,\mathcal{B}\,w_2)\,.
\end{multline}
After that, $|\Omega^{(2)\:\!i}\,{}_j|_{i=j}=0$. Unfortunately, it is not possible to find the metric functions made of the simple, elementary functions as these given above, for which all the other energy-momentum components are equal to zero simultaneously. In Appendix, we show the exact solution up to second order. One can observe, that the metric functions are much more complicated there. Nevertheless, after substitution (\ref{Eqn:Metric2}) alone, the remaining energy-momentum tensor elements in the second order are small at the time $\EdStime$: $\max\limits_{x^\mu\in\mathcal{D}}|\Omega^{(2)\:\!i}\,{}_j|_{i\neq j}=2.2\times 10^{-10}$ and $\max\limits_{x^\mu\in\mathcal{D}}|\Omega^{(2)\:\!0}\,{}_i|=1.5\times 10^{-6}$. To ensure that these values are negligible one can compare them with the third order contribution: $\max\limits_{x^\mu\in\mathcal{D}}|\Omega^{(3)\:\!i}\,{}_j|_{i=j}=2.2\times 10^{-3}$. Now, instead of trying to get the exact dust solution in the second order we can cancel out the maximal contribution from the third order.

\subsection{The third order}
The structure of the energy-momentum tensor elements in the third order is similar to that of the second order. In the same manner, by introducing the only nonzero metric functions:
\begin{multline}\label{Eqn:Metric3}
A^{(3)}(w)=t^{-3}\,(\EdStime)^3\,\left(\frac{\sin^4(\mathcal{B}\,w)}{16}-\frac{3\,\sin^2(\mathcal{B}\,w)}{64}+\frac{1}{96} \right)\,,\\
F^{(3)}(w_1,w_2)=\frac{t^{-11/3}\,(\EdStime)^3\,\mathcal{C}^2}{32\,\mathcal{B}^2}\,f(w_1,w_2)\,,\\
f(w_1,w_2)=\left(\sin^3(\mathcal{B}\,w_1)\sin(\mathcal{B}\,w_2)\cos(\mathcal{B}\,w_1)\cos(\mathcal{B}\,w_2)+\dots\right. \\
\dots+\sin(\mathcal{B}\,w_1)\sin^3(\mathcal{B}\,w_2)\cos(\mathcal{B}\,w_1)\cos(\mathcal{B}\,w_2)-\dots \\
\left.\dots-2\,\sin(\mathcal{B}\,w_1)\sin(\mathcal{B}\,w_2)\cos(\mathcal{B}\,w_1)\cos(\mathcal{B}\,w_2)\right)\,,
\end{multline}
one can obtain $|\Omega^{(3)\:\!i}\,{}_j|_{i=j}=0$. As previously, the remaining values are small: $\max\limits_{x^\mu\in\mathcal{D}}|\Omega^{(3)\:\!i}\,{}_j|_{i\neq j}=1.2\times 10^{-10}$ and $\max\limits_{x^\mu\in\mathcal{D}}|\Omega^{(3)\:\!0}\,{}_i|=8.5\times 10^{-7}$. The deviation from the dust energy-momentum tensor which comes from the second and third order is smaller than the maximal contribution from the fourth order: $\max\limits_{x^\mu\in\mathcal{D}}|\Omega^{(4)\:\!i}\,{}_j|_{i=j}=3.5\times 10^{-4}$. The procedure of approaching the dust energy-momentum tensor, with the help of the $A^{(k)}$ and $F^{(k)}$ functions, can be continued in the fourth and higher orders until one reaches the second order limit of around $10^{-6}$. However, the result which we get so far is good enough. The deviation of order $10^{-4}$ that we get at $\EdStime$ for the amplitude $\lambda=0.26$ is comparable to the second order deviation in the linear perturbation theory with the much smaller amplitude $\lambda\approx10^{-2}$. 

\begin{figure}[h]
	\centering
	\includegraphics[width=0.49\textwidth]{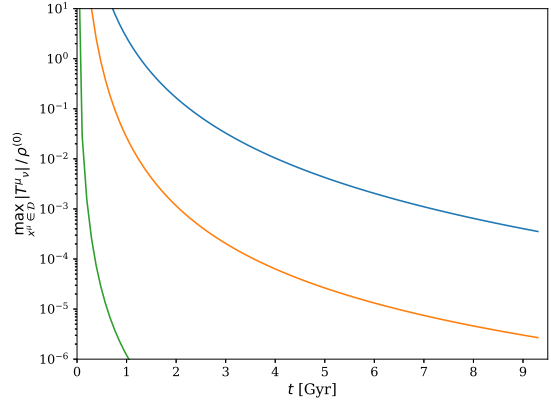}
	\caption{\label{fig:stresstensor} \scriptsize{The energy-momentum tensor elements as functions of time. The \emph{blue curve} represents $\max\limits_{x^\mu\in\mathcal{D}}|T^{(4)\,i}\,{}_j|_{i=j}/\rho^{(0)}$, the \emph{orange} - $\max\limits_{x^\mu\in\mathcal{D}}|T^{(2)\,0}\,{}_i+T^{(3)\,0}\,{}_i+T^{(4)\,0}\,{}_i|/\rho^{(0)}$, while the \emph{green} one - $\max\limits_{x^\mu\in\mathcal{D}}|T^{(2)\,i}\,{}_j+T^{(3)\,i}\,{}_j+T^{(4)\,i}\,{}_j|_{i\neq j}/\rho^{(0)}$.}}
\end{figure} 
The above estimation was made at the time $\EdStime$. Now, in Fig. \ref{fig:stresstensor}, we plot the time dependence of the energy-momentum tensor elements for the metric functions (\ref{Eqn:Metric1},\ref{Eqn:Metric2},\ref{Eqn:Metric3}). One can see that the approximation $T_{\mu\:\!\nu}(\lambda)\approx T_{\mu\:\!\nu}^{(dust)}$ is good for a late times $t>3\,\mathrm{Gyr}$. For the small $t$, the energy-momentum tensor elements other than the density become important and one cannot expect that the space-time metric describes the dust universe there. 

\begin{figure}[h]
	\centering
	\includegraphics[width=0.49\textwidth]{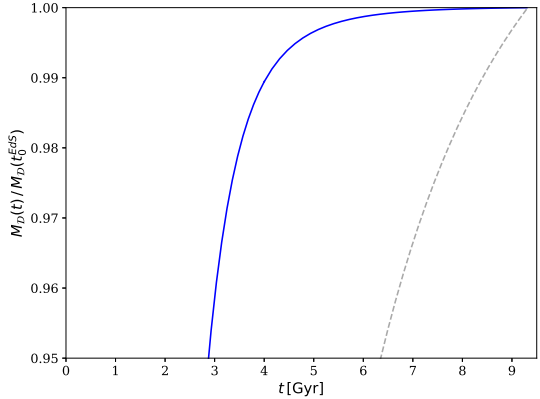}
	\caption{\label{fig:mass} \scriptsize{The normalized mass of the elementary cell as the function of time is plotted in \emph{blue}. For comparison, the \emph{gray dashed curve} relates to the metric restricted to the linear order with the same amplitude $\lambda=0.26$. }}
\end{figure}

For the universe filled with the dust, the total mass within the elementary cell $M_{\mathcal{D}}(t)$ should be conserved in time. (see for example \cite{2000GReGr..32..105B}). Once the metric is explicitly given, one can obtain $M_{\mathcal{D}}(t)$ by a direct numerical integration:
\begin{equation}
M_{\mathcal{D}}(t)=\int_{\mathcal{D}}\ud^3 x\,\sqrt{\mathrm{det}\,g_{i\:\!j}}\,\rho(t,\vec{x})\,.
\end{equation}
The result is plotted in Fig. \ref{fig:mass}. The mass discrepancy for the times $t>5\,\mathrm{Gyr}$ is lower than $0.5\%$. Therefore, we ensure ourselves that at that times our model approximates well the dust universe with a relatively high amplitude of the overdensities. Moreover, it is evident that the higher order terms in the metric are necessary to achieve this accuracy. In the next section, we will study the basic properties of the model.

\section{The model properties}\label{sec:properties}
\subsection{The density distribution}
\begin{figure}[h]
	\centering
	\includegraphics[width=0.5\textwidth]{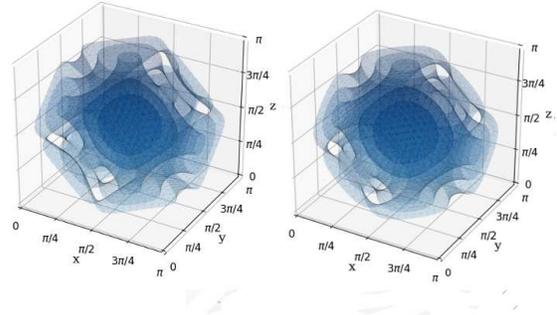}
	\caption{\label{fig:density} \scriptsize{\emph{Left:} the model isodensity surfaces within the elementary cell at the time $t=3\,\mathrm{Gyr}$. \emph{Right:} the same picture at the time $t=\EdStime=9.32\,\mathrm{Gyr}$. }}
\end{figure}

In the calculation of the total density, we will restrict to the fourth order $\rho=\sum\limits_{k=0}^{4}\,\lambda^k\,\rho^{(k)}$. The contribution $\rho^{(k)}=-T^{(k)\:\!0}\,{}_0$ behaves roughly as $\rho^{(k)}\propto (\EdStime)^k\,t^{-k-2}$. The higher orders are more important at early times and give less impact on the total density at the late times. The analytical formula for $\rho$ is the large, but relatively simple expression containing some products of the powers of the trigonometric functions. In Fig. \ref{fig:density}, we show the isodensity surfaces at the time $\EdStime$ and $t=3\,\mathrm{Gyr}$. Although the time dependence of $\rho^{(k)}$ in each order is different, the shape of the isodensity surfaces does not change much during the time evolution. Following \cite{2000GReGr..32..105B}, by introducing the volume of the elementary cell at the hypersurface of a constant time:
\begin{equation}\label{Eqn:volume}
V_{\mathcal{D}}=\int_{\mathcal{D}}\ud^3 x\,\sqrt{\mathrm{det}\,g_{i\:\!j}}\,,
\end{equation}
we may define the average density $\langle\rho\rangle_{\mathcal{D}}=M_\mathcal{D}/V_{\mathcal{D}}$. At the time $\EdStime$, the  resulting average density in the critical units is $\langle\Omega\rangle_{\mathcal{D}}=1.23$, where $\Omega=\rho/\rho_{cr}$. At that time, the observer located at the overdensity $\vec{x}_{\mathcal{O}}$ measures $\Omega=1.52$, while at the underdensity $\vec{x}_{\mathcal{U}}$ one gets $\Omega=0.97$.

\subsection{The Hubble parameter}
\begin{figure}[h]
	\centering
	\includegraphics[width=0.49\textwidth]{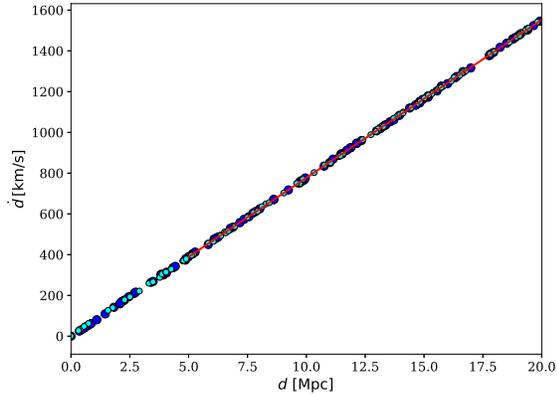}
	\caption{\label{fig:HubbleDiagram} \scriptsize{The Hubble diagram for the observers at the time $\EdStime$. The \emph{blue} points are generated for the observer located at the overdensity, while the \emph{cyan} points are plotted for the observer in the underdensity. The linear fit is depicted by the \emph{red line}.}}
\end{figure}

For the observer located at the time $\EdStime$ at $\vec{x}_{\mathcal{O}}$ or $\vec{x}_{\mathcal{U}}$ respectively, we performed the following numerical experiment. With the probability distribution uniform on the unit sphere, we generated ten random directions $(\theta,\phi)$. For each direction, we generated randomly ten points lying on the curve $\gamma(l)=(\EdStime,x_0+l\,\sin\theta\cos\phi, y_0+l\,\sin\theta\sin\phi, z_0+l\,\cos\theta)$, where $\vec{x}=(x_0,y_0,z_0)$ is the observer position. Then, by performing the numerical integration we obtain the distance $d(\widetilde{l})$ to each point:
\begin{equation}
d(\widetilde{l})=\int\limits_{0}^{\widetilde{l}}\sqrt{\gamma'(l)^i\,\gamma'(l)^j\,g_{i\:\! j}}\,\mathrm{d}l
\end{equation}
The integration kernel is the explicit function of time since the metric elements depend on $t$. By taking the time derivative of the integration kernel and calculating the numerical integral again, we obtain the velocity of each point $\dot{d}(\widetilde{l})$. This allows us to make the Hubble diagram presented in Fig \ref{fig:HubbleDiagram}. 
The linear fit to the points located further than $5\,\mathrm{Mpc}$ gives the value of the Hubble constant $H_0=77.68\,\mathrm{km/s/Mpc}$ for the observer at $\vec{x}_{\mathcal{O}}$, and $H_0=77.42\,\mathrm{km/s/Mpc}$ for the observer located at $\vec{x}_{\mathcal{U}}$. In the considered space-time there are no significant differences in the values of the Hubble constant between the overdense and underdense regions. However, if we demand that the observer should measure the Hubble constant comparable to the real value, we should locate him in another time position.

\begin{figure}[h]
	\centering
	\includegraphics[width=0.49\textwidth]{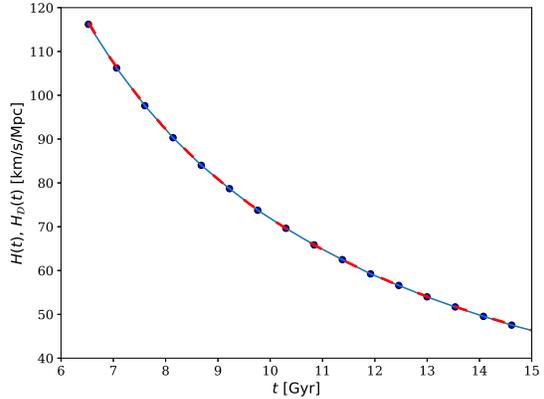}
	\caption{\label{fig:HubbleParameter} \scriptsize{The plot of the Hubble parameter $H(t)$. The \emph{blue points} are generated with the method described in the text. \emph{Light blue curve} is the cubic spline interpolation of these points. The effective Hubble parameter $H_{\mathcal{D}}(t)$ is shown by the \emph{red dashed curve}.}}
\end{figure}

To find the time coordinate $\univage$, in which the Hubble constant is equal to $H_0=70.0\,\mathrm{km/s/Mpc}$, we repeat the procedure of the Hubble diagram construction placing the observer in the overdensity $\vec{x}_{\mathcal{O}}$ at a discrete set of the time coordinate values. In each case, we perform the linear fit to the generated data $(d,\dot{d})$ and in effect, we get a discrete set of points in a Hubble parameter $H(t)$ plot. We show them as the blue points in Fig \ref{fig:HubbleParameter}. After that, we use the cubic spline interpolation shown as the light blue curve. The resulting value of the universe age for which the $H_0=70.0\,\mathrm{km/s/Mpc}$ is $\univage=10.25\,\mathrm{Gyr}$. 

Sometimes it is convenient to use the domain dependent effective Hubble parameter $H_{\mathcal{D}}$, defined as in \cite{2000GReGr..32..105B}. We will consider the elementary cell as the domain $\mathcal{D}$. For the selected domain, one can introduce the effective scale factor $a_{\mathcal{D}}(t)=V^{1/3}_{\mathcal{D}}(t)/V^{1/3}_{\mathcal{D}}(\univage)$. Then, by definition $H_{\mathcal{D}}(t)=\dot{a}_{\mathcal{D}}(t)/a_{\mathcal{D}}(t)$. Numerically more efficient is to take the time derivative of the integration kernel of (\ref{Eqn:volume}), perform the numerical integration again to obtain $\dot{V}_{\mathcal{D}}$ and calculate the effective Hubble parameter from $H_{\mathcal{D}}=\dot{V}_{\mathcal{D}}(t)/(3\,V_{\mathcal{D}}(t))$. In Fig. \ref{fig:HubbleParameter}, the resulting $H_{\mathcal{D}}(t)$ is plotted by the red dashed curve. One can see that it is consistent with the Hubble parameter $H(t)$ calculated previously. We want to emphasize that we computed both quantities in a straightforward way, directly from the space-time metric.

At the end of this paragraph, we want to analyze the model density at the time $\univage$. The maximum density at $\vec{x}_{\mathcal{O}}$ is $\Omega=1.22$. The minimum density for the observer located at $\vec{x}_{\mathcal{U}}$ is $\Omega=0.82$. The average over the elementary cell is $\langle\Omega\rangle_{\mathcal{D}}=1.003$. It seems that the average density at the time $\univage$, in which the observer measures the Hubble constant $H_0$, is close to the EdS model with the same $H_0$ value. Below, we discuss this issue in details.

\subsection{The EdS as the model average}\label{sec:EdS}
When one considers the limit $\lambda\to 0$, the metric tends to the background metric $g^{(0)}$, which is the EdS space-time. On the other hand, because we used a quite high amplitude $\lambda=4/15$, the metric $g$ is not close to $g^{(0)}$. This situation is qualitatively different from the widely used Swiss cheese models, where the space-time regions describing inhomogeneities are glued with the background FLRW model. Here, there are no space-time regions with the FLRW symmetry and the inhomogeneities cover the entire space. It is then interesting that the EdS model is a possible candidate for the model average.

In the FLRW model, one can introduce the parameter $\Omega_k=-k/(H^2(t)\,a^2(t))=-R/(6\,H^2(t))$, which is the measure of the curvature of space. In this formula, $k$ is the curvature parameter from the FLRW metric and $R$ is the scalar curvature of the hypersurface of the constant time. In analogy, one can define the averaged curvature parameter at the time $\univage$: $\langle\Omega_R\rangle_{\mathcal{D}}=-\langle R\rangle_{\mathcal{D}}/(6\,H_0^2)$. In the presented model it has the value $\langle\Omega_R\rangle_{\mathcal{D}}=-8.9\times10^{-5}$, which means that the space is almost flat. 

\begin{figure}[h]
	\centering
	\includegraphics[width=0.49\textwidth]{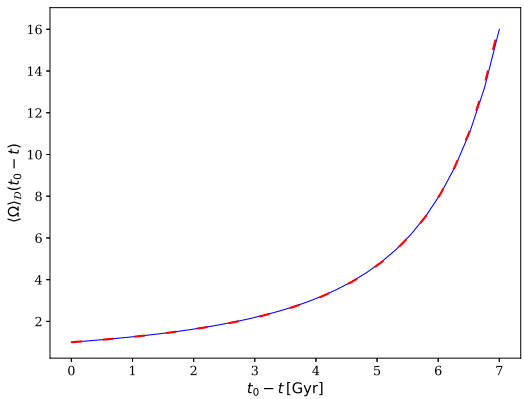}
	\caption{\label{fig:density2} \scriptsize{The \emph{blue curve} - the time dependence of the averaged density of the model, with respect to the time $t_0=\univage$. The \emph{dashed red curve} - the EdS prediction calculated with respect to the time $t_0=\EdStime$.}}
\end{figure}

\begin{figure}[h]
	\centering
	\includegraphics[width=0.49\textwidth]{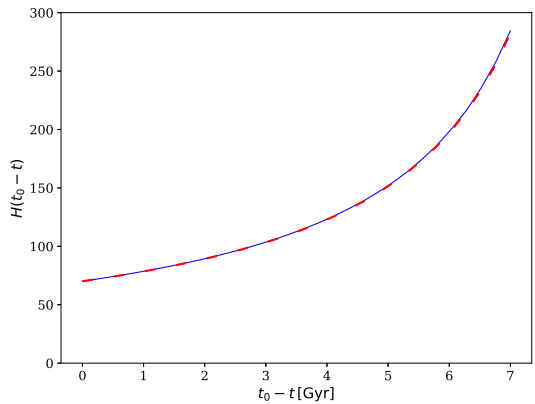}
	\caption{\label{fig:Hubble2} \scriptsize{The \emph{blue curve} - the time dependence of the model Hubble parameter, with respect to the time $t_0=\univage$. The \emph{dashed red curve} - the EdS prediction calculated with respect to the time $t_0=\EdStime$.}}
\end{figure}

In Fig. \ref{fig:density2} we present the time evolution of the model density, while in Fig. \ref{fig:Hubble2} we show the time evolution of the Hubble parameter. To be able to compare them with the EdS prediction, both figures were plotted with respect to the time $t_0$, in which the observer measures the value of the Hubble constant $H_0=70.0\,\mathrm{km/s/Mpc}$. As it is seen, there is a perfect agreement between the model behavior and the EdS prediction. Therefore we conclude, that the EdS space-time describes the average density and the average expansion very well.

\subsection{The light propagation}
The procedure of the angular diameter distance-redshift relation derivation is similar to that in the paper \cite{2017PhRvD..95f3517S}. We will locate the observer in the three different positions: in the overdensity $\vec{x}_{\mathcal{O}}=(\frac{\pi}{2},\frac{\pi}{2},\frac{\pi}{2})$, in the underdensity $\vec{x}_{\mathcal{U}}=(0,0,0)$ and at some point somewhere in between $\vec{x}_{\mathcal{M}}=(0.7,1.1,2.1)$, and in the two time instants $\univage$ and $\EdStime$. The density at the middle position $\vec{x}_{\mathcal{M}}$ at the time $\univage$ is $\Omega=1.067$, so it is the point with a slightly higher density than the average. 

For each observer position, we generate one hundred random directions $\vec{k}$, with a probability distribution uniform on the unit sphere. Since $|\vec{k}|=1$, we choose $k^0$ so that $k^\mu\,k_\mu=0$ at the observer position and $k^0<0$, so the geodesic is past oriented. With the observer position $x^\mu$ and wave vector $k^\mu$ as the initial conditions, we solve with the help of the fourth order Runge-Kutta method the geodesic equation written in the form of the two first-order differential equations:
\begin{equation}
\frac{\ud k^\mu}{\ud l}=-\Gamma^\mu_{\alpha\:\!\beta}\,k^\alpha\,k^\beta\,,
\end{equation}
\begin{equation}
\frac{\ud x^\mu}{\ud l}=k^\mu\,.
\end{equation}
To check the numerical accuracy we generate the reference geodesic, for which we correct $|\vec{k}|$ in each Runge-Kutta step to satisfy $k^\mu\,k_\mu=0$ exactly. Then we choose so small Runge-Kutta step $\Delta l=0.05$ so that the difference between the geodesics generated by both methods is negligible.

After we get the resulting geodesic $x^\mu(l)$, we obtain the redshift along it from the definitions $z=(\omega_{em}-\omega_{obs})/\omega_{obs}$  and $\omega=U^\mu\,k_\mu$. We assume that the light emitter and the observer are comoving with matter, so their four-velocity is represented by the vector $U^\mu=(1,0,0,0)$, while $k^\mu(l)$ is the wave vector along the geodesic. 

To derive the angular diameter distance $d_A$ along the geodesic we use the Sachs formalism \cite{1961RSPSA.264..309S}. We construct the Sachs basis $(s_1^\mu,s_2^\mu)$ at the observer position. The Sachs basis vectors are orthogonal to each other and orthogonal to the wave vector $k^\mu$ and the observer four-velocity $U^\mu$ respectively. Next, for each of the Sachs basis vector, we solve numerically the equation of the parallel transport along the geodesic. This way, the Sachs basis is defined at each point of the geodesic. Then, to calculate the angular diameter distance along the geodesic we solve with the help of the fourth order Runge-Kutta method the following system of equations. The focusing equation rewritten as the two first order differential equations:
\begin{equation}\label{eqn:Focusing}
\frac{\ud}{\ud  l}\,\dot{d}_A=-\left(\frac{1}{2}\,R_{\mu\:\!\nu}\,k^\mu\,k^\nu+|\sigma|^2 \right)\,d_A\,,
\end{equation}
\begin{equation}
\frac{\ud}{\ud l}d_A=\dot{d}_A\,,
\end{equation}
and the Sachs evolution equations for the components of the complex shear $\sigma$:
\begin{equation}\label{eqn:Sachs_sigma}
\frac{\ud}{\ud l}\sigma_1+2\,\sigma_1\,\theta=-\frac{1}{2}\,C_{\alpha\:\!\beta\:\!\gamma\:\!\delta}\,\left(s_1^\alpha\,k^\beta\,k^\gamma\,s_1^\delta+s_2^\alpha\,k^\beta\,k^\gamma\,s_2^\delta \right)\,,\nonumber
\end{equation}
\begin{equation}
\frac{\ud}{\ud l}\sigma_2+2\,\sigma_2\,\theta=C_{\alpha\:\!\beta\:\!\gamma\:\!\delta}\,s_1^\alpha\,k^\beta\,k^\gamma\,s_2^\delta\,,
\end{equation}
where $C_{\alpha\:\!\beta\:\!\gamma\:\!\delta}	$ is the Weyl tensor and the scalar expansion rate $\theta$ is substituted by $\theta=\dot{d}_A/d_A$. Here, by the dot we mean the derivative over the affine parameter along the geodesic $\dot{}\equiv\ud/\ud l$. The initial conditions at $l=0$ are: $d_A=10^{-6}\,\mathrm{Mpc}$, $\dot{d}_A=1$, $\sigma=0$. For numerical reasons, we start with the $d_A$ slightly above zero, so that the expansion scalar is finite at $l=0$. As the result of the presented procedure, we obtain the angular diameter distance along each geodesic $d_A(l)$. Finally, by using the information about the redshift along the geodesic we get the angular diameter distance as a function of redshift $d_A(z)$. At the end, we checked the convergence of the resulting $d_A(z)$ with a decreasing Runge-Kutta step.

\begin{figure}
	\centering
	\includegraphics[width=0.49\textwidth]{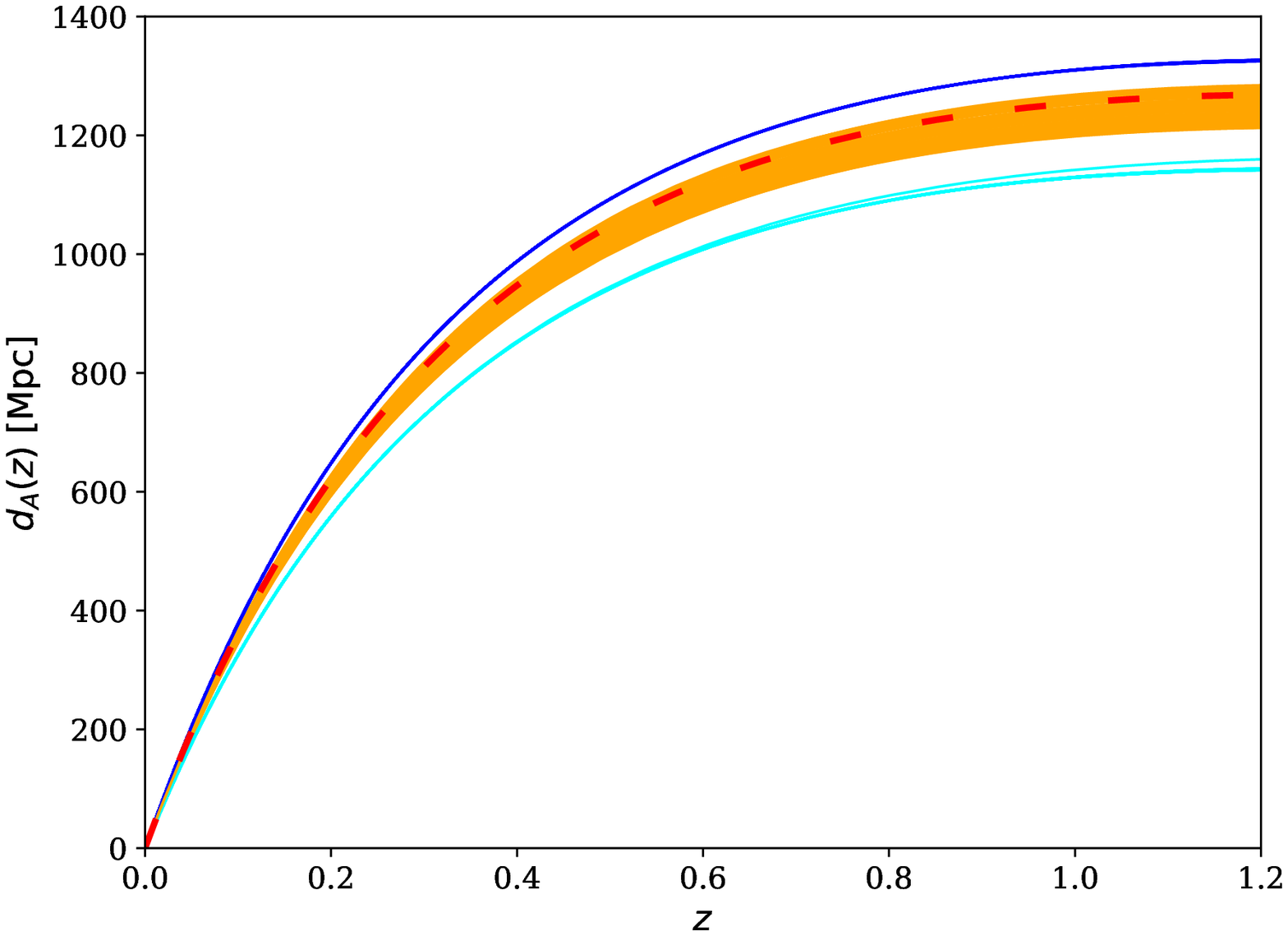}
	\includegraphics[width=0.49\textwidth]{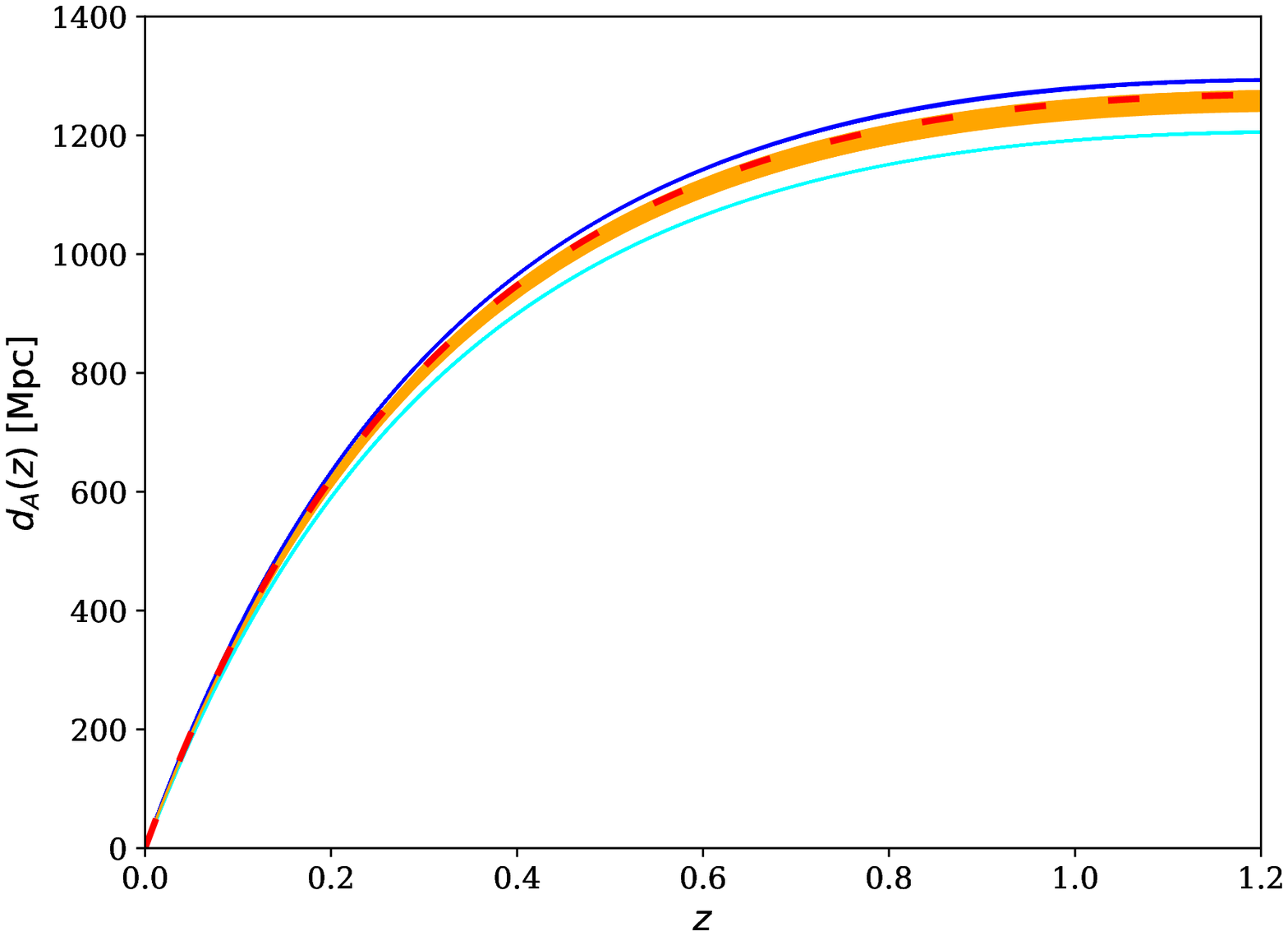}
	\includegraphics[width=0.49\textwidth]{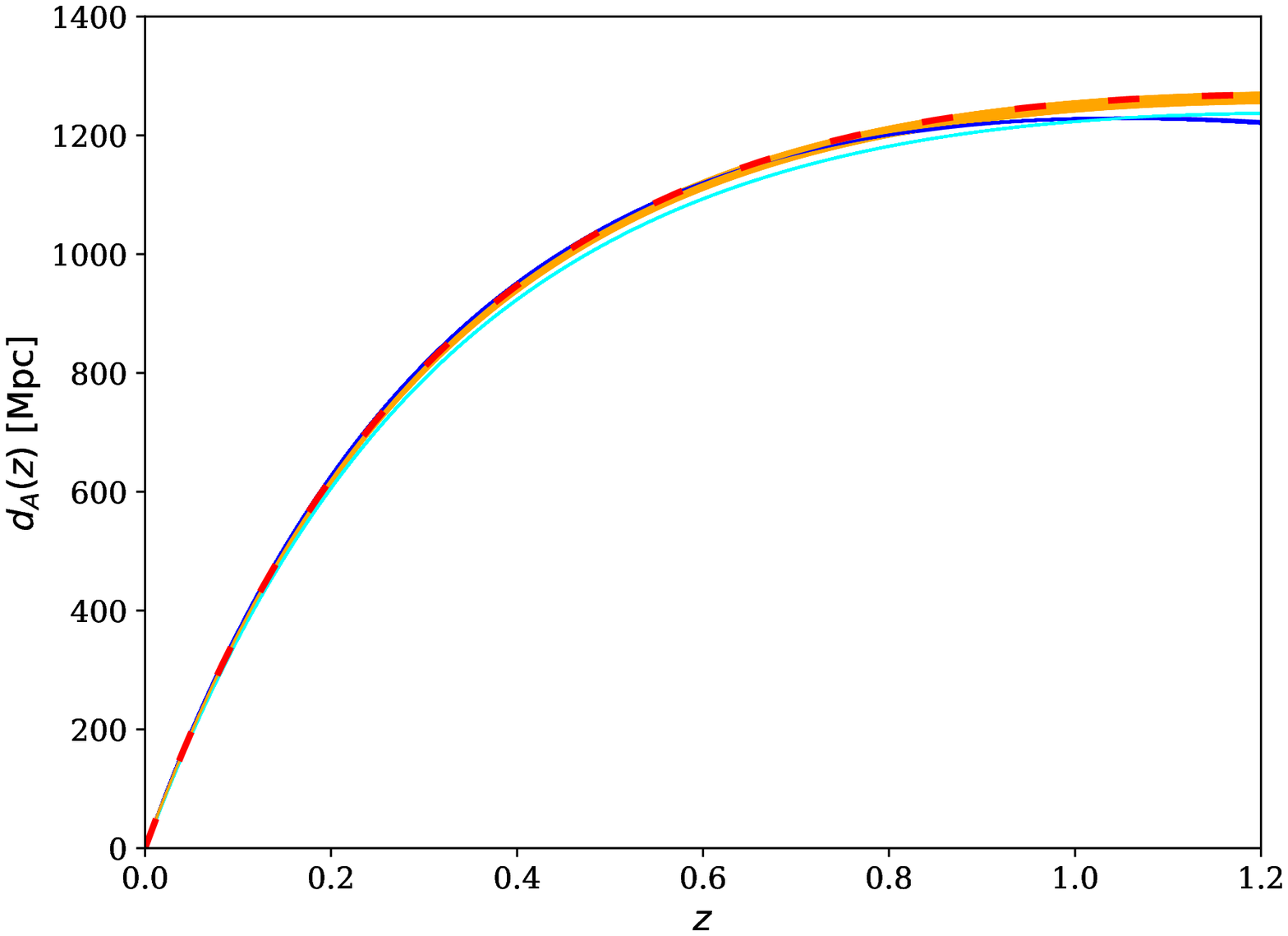}
	\caption{\label{fig:results1} \scriptsize{The angular diameter distance-redshift relation $d_A(z)$. The observer is located at the time $\EdStime$ at the overdensity $\vec{x}_{\mathcal{O}}=(\frac{\pi}{2},\frac{\pi}{2},\frac{\pi}{2})$ (\emph{blue}), at the underdensity $\vec{x}_{\mathcal{U}}=(0,0,0)$ (\emph{cyan}) or in the middle $\vec{x}_{\mathcal{M}}=(0.7,1.1,2.1)$ (\emph{orange}). Top panel corresponds to the amplitude $\lambda=4/15$, the middle panel - $\lambda=2/15$ and the bottom panel - $\lambda=1/15$. \emph{Red dashed curve} - EdS prediction.}}
\end{figure}
\begin{figure}
	\centering
	\includegraphics[width=0.49\textwidth]{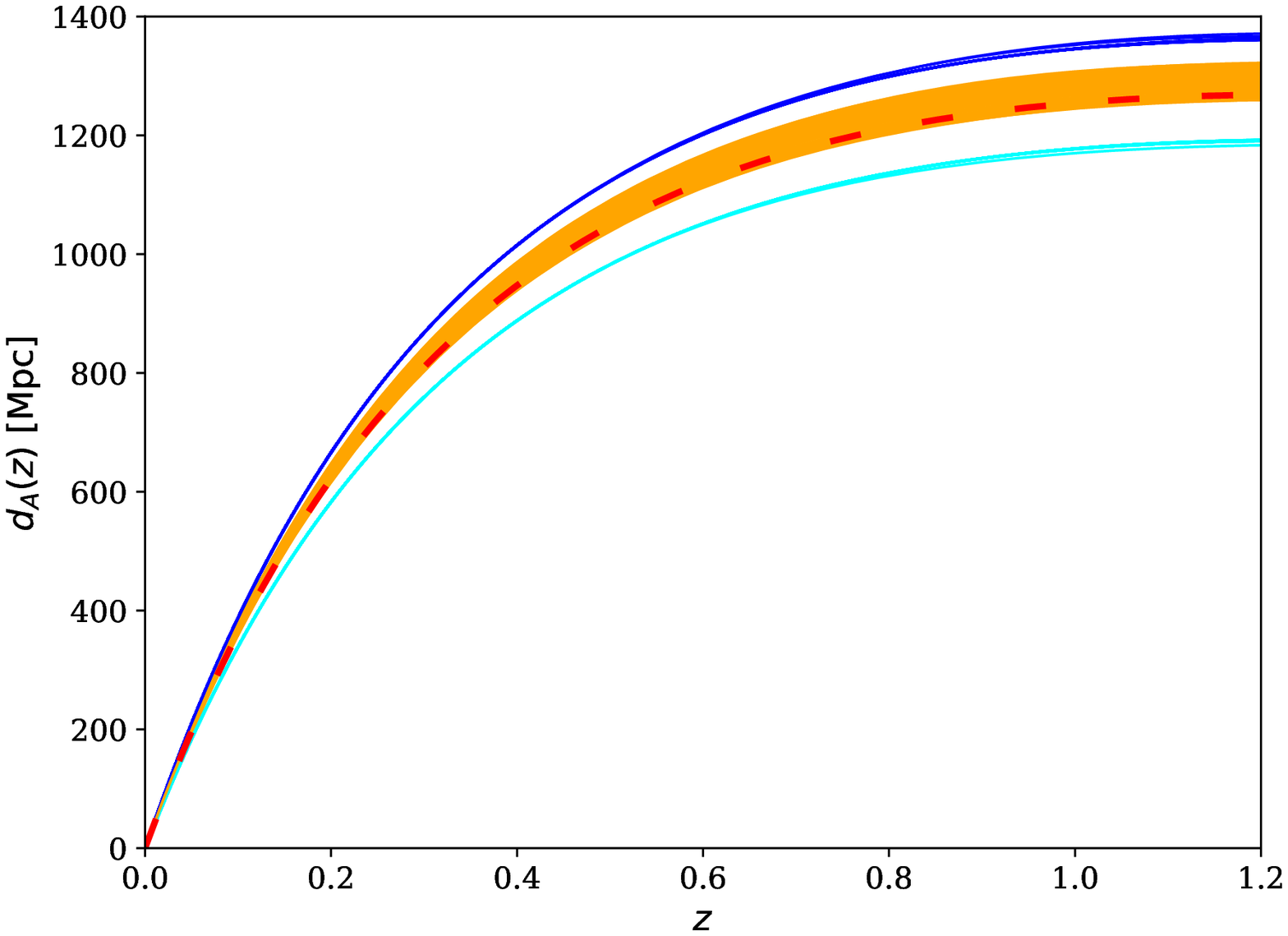}
	\includegraphics[width=0.49\textwidth]{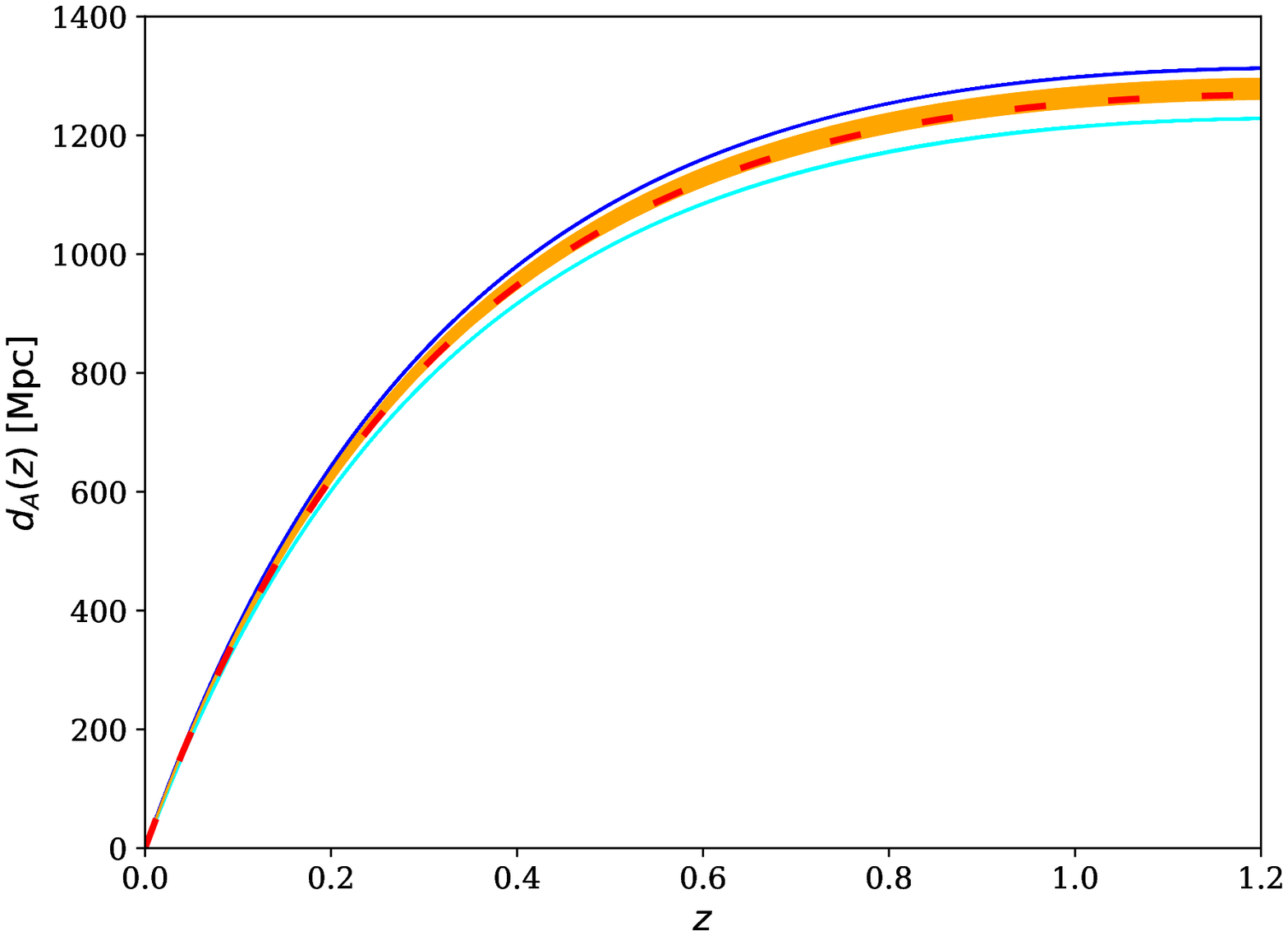}
	\includegraphics[width=0.49\textwidth]{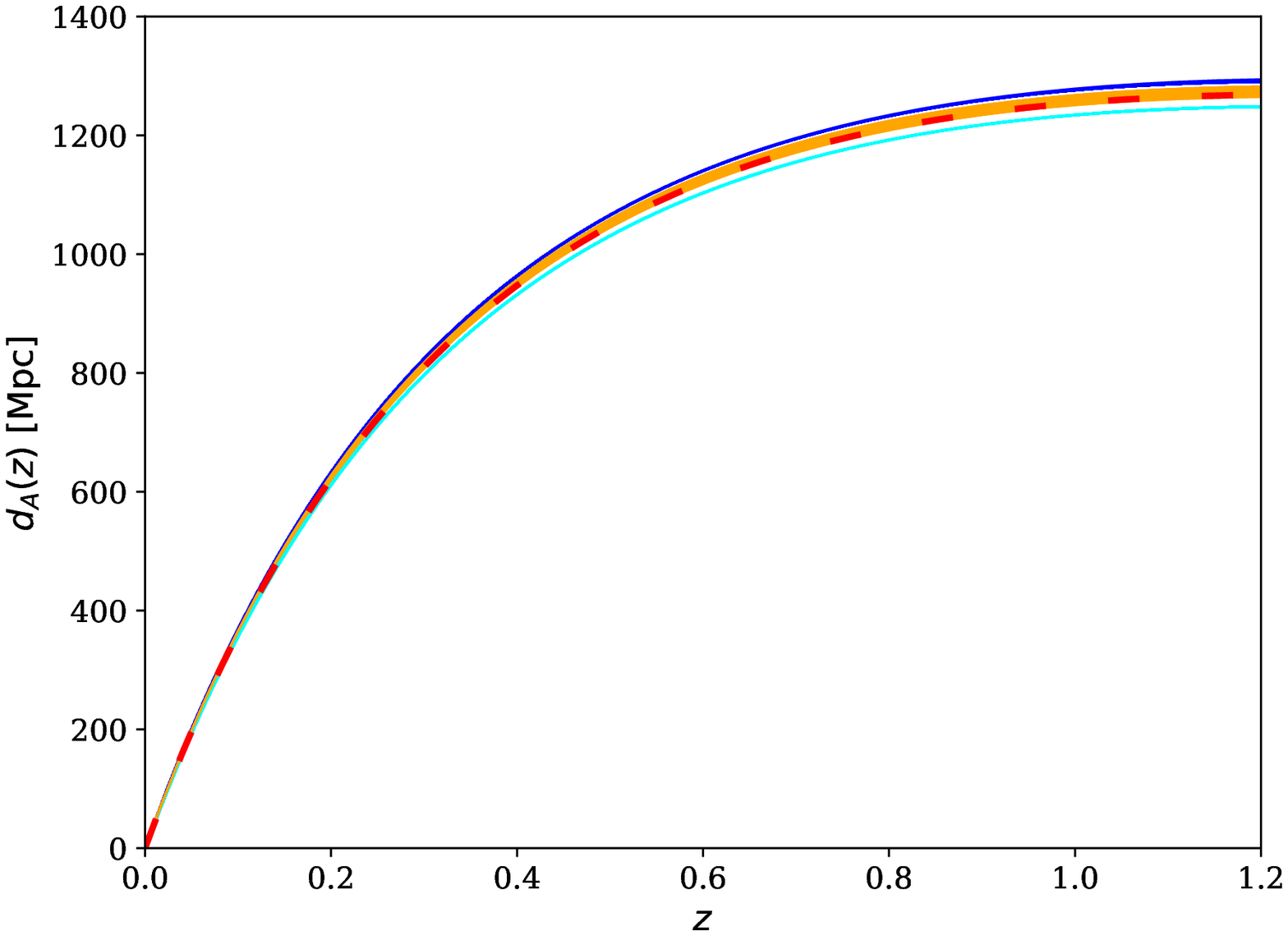}
	\caption{\label{fig:results2} \scriptsize{The similar plots of the $d_A(z)$ as in the Fig. \ref{fig:results1}, although for the observer located at the time $\univage$.}}
\end{figure}
We present the results in Figures \ref{fig:results1} and \ref{fig:results2}. In each figure, the set of the blue curves shows the $d_A(z)$ for the one hundred geodesics generated for the observer at the overdensity $\vec{x}_{\mathcal{O}}$, the similar set of the cyan curves correspond to the observer at the underdensity $\vec{x}_{\mathcal{U}}$ while the orange curves refer to the observer at the middle position $\vec{x}_{\mathcal{M}}$. The red dashed line is the EdS prediction. The plots are prepared for the following values of the amplitude $\lambda\in\{4/15,2/15,1/15\}$ and for the observer located in the two time instants $t\in\{\EdStime,\univage\}$. 

On basis of these results we conclude that although the EdS space-time describes well the average properties of the matter distribution and expansion rate of the considered model, the light propagation differs significantly from the EdS prediction. The behavior of the $d_A(z)$ depends on the observer position. For the observer located in the point with the local density lower than the average, the resulting $d_A(z)$ is lower than the EdS reference curve. If the observer's local density is higher than the average, then the resulting $d_A(z)$ is larger than the EdS curve. The deviation from the EdS is the greatest for the observer located in the maximum or the minimum of the density distribution. The greater the amplitude $\lambda$, the larger the differences between different curves. If one takes the small amplitude, locates the observer at the time $\EdStime$ and restricts the metric to the linear order, then the resulting $d_A(z)$ tends to the curves we have shown in our previous paper, where we do not observe the position dependence of the results.

\begin{figure}[h]
	\centering
	\includegraphics[width=0.49\textwidth]{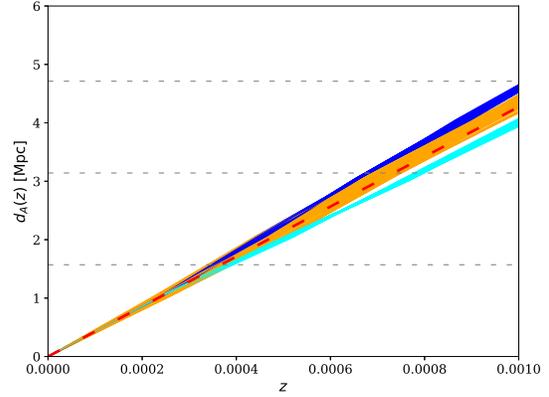}
	\caption{\label{fig:results3} \scriptsize{The angular diameter distance-redshift relation $d_A(z)$ for the observers at $\univage$. The low redshift range. Colors as in the Fig. \ref{fig:results1}.}}
\end{figure}
Another interesting property is the width of the bundle of the $d_A(z)$ curves. The bundle of one hundred geodesics generated from the middle position $\vec{x}_{\mathcal{M}}$ characterizes by the larger width in the angular diameter distance-redshift relation than the bundle related to the overdensity or the underdensity. In the positions $\vec{x}_{\mathcal{O}}$, $\vec{x}_{\mathcal{U}}$ the density distribution is much more symmetrical than in the middle $\vec{x}_{\mathcal{M}}$. This could indicate that the local neighborhood of the observer plays the crucial role here. To confirm this intuition we plot in Fig. \ref{fig:results3} the $d_A(z)$ relation for the low redshifts. One can see that the separation between the curves begin around $d_A\approx 1\,\mathrm{Mpc}$. For the twenty curves $\gamma(p)=(\univage,p\,\sin\theta\cos\phi,p\,\sin\theta\sin\phi,p\,\cos\theta)$, where $(\theta,\phi)$ are the random directions, the average length from $p=0$ to $p=\pi/2$ is $d=1.605\,\mathrm{Mpc}$. The same calculation for the curves centered at the overdensity gives $d=1.538\,\mathrm{Mpc}$. Since $\pi$ is the coordinate size of the elementary cell, in view of these estimations, the value around $1.57\,\mathrm{Mpc}$ can be thought as the physical radius of the inhomogeneities at $\univage$. The separation of the $d_A(z)$ curves takes place while the
light ray passes through the observer neighborhood to
the nearest region with a different density.

\begin{figure}[h]
	\centering
	\includegraphics[width=0.49\textwidth]{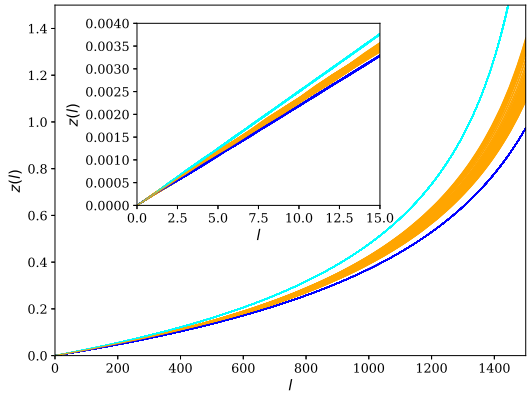}
	\caption{\label{fig:results4} \scriptsize{The redshift as a function of the affine parameter along the geodesic $z(l)$ for the observers located at the time $\univage$. \emph{Blue} - observer in the overdensity $\vec{x}_{\mathcal{O}}$, \emph{cyan} - observer in the underdensity $\vec{x}_{\mathcal{U}}$, \emph{orange} - the middle position of the observer $\vec{x}_{\mathcal{M}}$.}}
\end{figure}

As we have seen in section \ref{sec:EdS}, the space is almost flat. This means that the Ricci term in the focusing equation (\ref{eqn:Focusing}) is very small. Since initial shear is equal to zero, for small distances the right-hand side of the focusing equation is close to zero. Therefore, the angular diameter distance as a function of the affine parameter $d_A(l)$ is almost linear there. This suggests that the separation of the $d_A(z)$ curves is caused by the local changes of redshift along the geodesics initiated in different environments. To show that, we plot in Fig. \ref{fig:results4} the relation between the redshift and the affine parameter along the resulting geodesics. For the local universe $d_A\approx l$. The separation of the curves around $l\approx 1.5$ is then consistent with that in Fig \ref{fig:results3}.

\begin{figure}[h]
	\centering
	\includegraphics[width=0.49\textwidth]{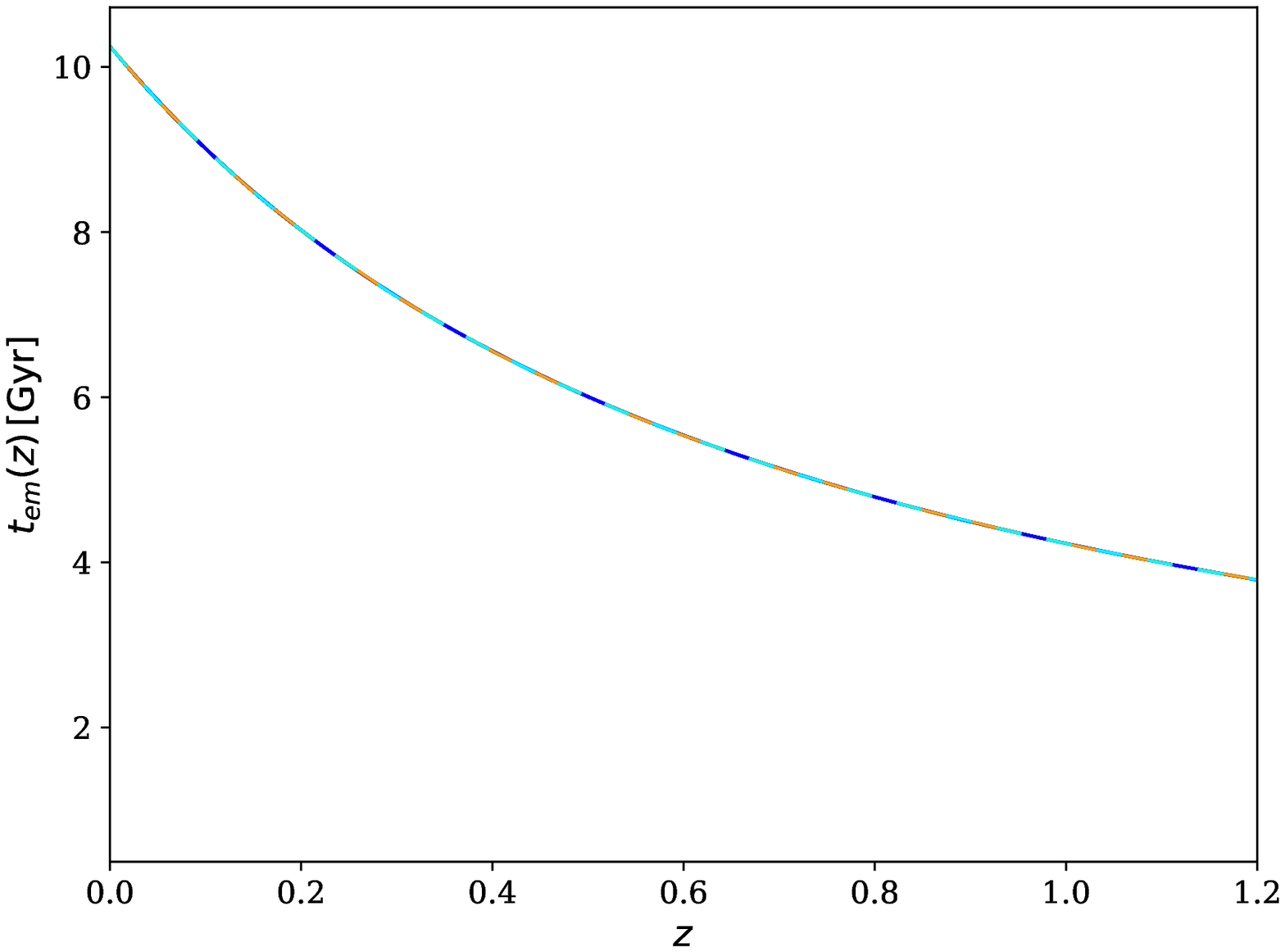}
	\caption{\label{fig:results5} \scriptsize{The time of the light emission as a function of redshift $t_{em}(z)$. The observers are located at the time $\univage$. \emph{Dashed blue} - observer in the overdensity $\vec{x}_{\mathcal{O}}$, \emph{dashed cyan} - observer in the undersity $\vec{x}_{\mathcal{U}}$, \emph{dashed orange} - the middle position of the observer $\vec{x}_{\mathcal{M}}$. }}
\end{figure}

In the end, it is instructive to show the emission time of the light as a function of redshift. This relation is plotted in Fig. \ref{fig:results5}. For a given redshift the emission time is the same, no matter where the observer is located. This is not surprising since the redshift $z$ is related to the time-like component of the wave vector $k^0$. One can connect the emission time $t_{em}$ with the effective scale factor at that time $a_{\mathcal{D}}(t_{em})$. Because the EdS model describes well the average expansion, the effective scale factor expressed as a function of redshift has the same form as in the EdS space-time:
\begin{equation}
a_{\mathcal{D}}(z)=\frac{1}{1+z}
\end{equation}

\section{Conclusions}
In the present paper, we constructed perturbatively the approximate model of the inhomogeneous universe for which the dust inhomogeneities are distributed in the infinite, periodic lattice on the Einstein-de Sitter background. This way we extend our previous work beyond the linear perturbation theory so that the larger amplitude of the inhomogeneities is allowed. We analyzed basic properties of the model. We have shown that the Einstein-de Sitter space-time describes well the time evolution of the model average density and the averaged expansion characterized by the Hubble parameter $H(t)$. On the other hand, the light propagation differs significantly from that in the EdS background. The angular diameter distance-redshift relation $d_A(z)$ is influenced by the inhomogeneities and depends on the observer's position.

In literature, the position dependence of $d_A(z)$ appears in some inhomogeneous models. In some of them, \mbox{e. g.} \cite{2007PhRvD..76l3004M}, the presence of the inhomogeneities could partly mimic the effect of the dark energy driven accelerated expansion. However, as it was pointed out in \cite{2006PhRvD..73h3519A}, the cosmological model which tries to avoid the cosmological constant should explain not only the $d_A(z)$ relation but also the variety of the other available observations, in particular, the isotropy of the CMB power spectrum. Nevertheless, if such effects like the strong position dependence of the $d_A(z)$ results are present in simple, inhomogeneous models with a reliable matter content, one cannot exclude that they appear also in the real Universe.

The model presented in the current paper has no ambition to describe the real Universe. It should be rather considered as a training model. There are some aspects of this model which should be improved first: \emph{(i)} In the current model, the amplitude of the inhomogeneities decreases with time. It is reasonable to look for a model with an increasing amplitude of the inhomogeneities with a controlled growth rate. \emph{(ii)} For the presented model, the approximation to the dust energy-momentum tensor holds for the late times only. It is necessary to construct the model which is valid also for the early universe, to be able to consider the CMB observations. \emph{(iii)} Some generalizations with a background space-time other than the Einstein-de Sitter will be useful.

Anyway, the presented framework is a step towards the more realistic inhomogeneous cosmological model constructed beyond the linear perturbation theory. We emphasize that the presented solution is not in the widely used Swiss cheese class of models, so it provides new possibilities. We also think that the models which offer the metric explicitly given, as our model does, are important because enable one to calculate the observables directly from the metric without additional simplifying assumptions.

\section*{Acknowledgements}
Publication supported by the John Templeton Foundation Grant \emph{Conceptual Problems in Unification Theories} (No. 60671).

\bibliography{SikoraGlod2018}
\bibliographystyle{unsrt}


\section*{Appendix: Strict metric of the perturbed model up to second order}

We construct a~sample of a~linearly perturbed spatially flat Friedmann--Lema\^{i}tre cosmological model with irrotational dust-like inhomogeneities up to second order. We consider a~cosmic fluid which is irrotational, nonconductive, inviscid and for which the spatial gradient of the pressure vanishes. For simplicity, we additionally assume that at the first order the magnetic part of the Weyl tensor vanishes which enables us to treat only the scalar perturbations and disables vector and tensor perturbations. Further, we assume that still at the first order the Ricci scalar of the three-spaces orthogonal to the fluid flow is zero which enables only the decreasing mode of perturbations and disables the growing one.

The assumed conditions allow the synchronous comoving gauge at both orders in the problem without loss of generality. The solution for the metric field of the space-time $g_{\mu\nu}$ and the velocity field of the fluid flow $u_{\nu}$ of the considered perturbed model take the form
\begin{align}
g_{\mu\nu}&=g^{(0)}_{\mu\nu}+\lambda g^{(1)}_{\mu\nu}+\frac{\lambda^2}{2}g^{(2)}_{\mu\nu},\\
g^{(0)}_{\mu\nu}&=\mathop{\mathrm{diag}}(-b^2,a^2,a^2,a^2),\\
u_{\nu}&=(-b,0,0,0),
\end{align}
where $a$ and $b$ are functions of the time coordinate $t$ and $\lambda$ is some small parameter. The first order correction to the metric has the following form
\begin{equation}
g^{(1)}_{\mu\nu}=2a^2\begin{pmatrix} 0& 0& 0& 0\\ 0& \partial_{xx}o& \partial_{xy}o& \partial_{zx}o\\ 0& \partial_{xy}o& \partial_{yy}o& \partial_{yz}o\\ 0& \partial_{zx}o& \partial_{yz}o& \partial_{zz}o\end{pmatrix}.
\end{equation}
Here, $o$ is a function of time and spatial coordinates $x$, $y$, $z$ which satisfies the equation
\begin{equation}
\partial_{t}o-\frac{b}{a^3}i=0,
\end{equation}
where $i$ is an arbitrary function of spatial coordinates. The second order correction is given as
\begin{equation}
g^{(2)}_{\mu\nu}=2a^2\begin{pmatrix} 0& 0& 0& 0\\ 0& c_{11}& c_{12}& c_{31}\\ 0& c_{12}& c_{22}& c_{23}\\ 0& c_{31}& c_{23}& c_{33}\end{pmatrix},
\end{equation}
where $c_{mn}$ are functions of time and spatial coordinates. Because of complexity of the equations at the second order, the functions $c_{mn}$ are determined with the function $i$ explicitly specified as
\begin{equation}
i=l_x+l_y+l_z.
\end{equation}
We used the abbreviated notation $l_x\equiv l(x)$. The function $l$ is assumed to satisfy the equation
\begin{equation}
\partial_{xx}l_x+\alpha^2l_x=0,
\end{equation}
where $\alpha$ is a constant thus $l$ is a~linear combination of sine and cosine functions. 

We get the following result
\begin{align}
c_{23}&=o_1l'_yl'_z,\\
c_{31}&=o_2l'_zl'_x,\\
c_{12}&=o_3l'_xl'_y,\\
c_{11}&=-\alpha^2v_1l_x^2-\frac{\alpha^2}{2}j_1l_yl_z-\alpha^2(o_2-\frac{1}{2}j_2)l_zl_x\notag\\
&-\alpha^2(o_3-\frac{1}{2}j_3)l_xl_y,\\
c_{22}&=-\alpha^2v_2l_y^2-\alpha^2(o_1-\frac{1}{2}j_1)l_yl_z-\frac{\alpha^2}{2}j_2l_zl_x\notag\\
&-\alpha^2(o_3-\frac{1}{2}j_3)l_xl_y,\\
c_{33}&=-\alpha^2v_3l_z^2-\alpha^2(o_1-\frac{1}{2}j_1)l_yl_z-\alpha^2(o_2-\frac{1}{2}j_2)l_zl_x\notag\\
&-\frac{\alpha^2}{2}j_3l_xl_y,
\end{align}
where $l'_x\equiv\partial_{x}l_x$ and $v_n$, $o_n$, $i_n$, $j_n$ are functions of time which satisfy the following equations
\begin{align}
\partial_{tt}v_n-\frac{\partial_{t}\frac{b}{a^3}}{\frac{b}{a^3}}\partial_{t}v_n+2\alpha^2\frac{b^2}{a^6}&=0,\\
\partial_{t}o_n-\frac{\alpha^2}{2}\frac{b}{a^3}i_n&=0,\\
\partial_{t}i_n+abj_n&=0,\\
\partial_{tt}j_n-\frac{\partial_{t}\frac{b}{a^3}}{\frac{b}{a^3}}\partial_{t}j_n+2\alpha^2\frac{b^2}{a^2}j_n+2\alpha^2\frac{b^2}{a^6}&=0.
\end{align}
For the considered model, the above solution for the metric functions is the simplest possible in a~sense that neither of the functions $v_n$, $o_n$, $i_n$, $j_n$ can be taken as null. The function $v_n$ is easy to find since
\begin{equation}
v_n=-\alpha^2c^2+\beta c+\gamma,
\end{equation}
where the auxiliary function $c$ satisfies
\begin{equation}
\partial_{t}c-\frac{b}{a^3}=0,
\end{equation}
and $\beta$, $\gamma$ are constants of integration. When the scale factor $a$ is a power function of time, then also $v_n$ is so. However, this is not the case for the function $j_n$ which is then a~combination of sine and cosine integral functions. This causes the metric to be a~highly complicated function of time.

The energy density $\rho$ and the pressure $p$ in the considered model are expressed by the metric functions as follows
\begin{align}
8\pi\rho&=3\frac{(\partial_{t}a)^2}{a^2b^2}\notag\\
&+\lambda\Bigl(-2\alpha^2\frac{\partial_{t}a}{a^4b}l_x-2\alpha^2\frac{\partial_{t}a}{a^4b}l_y-2\alpha^2\frac{\partial_{t}a}{a^4b}l_z\Bigr)\notag\\
&+\frac{\lambda^2}{2}\Bigl(2\alpha^2\frac{\partial_{t}a}{ab^2}\partial_{t}(-v_1-2\alpha^2c^2)l_x^2\notag\\
&\quad+2\alpha^2\frac{\partial_{t}a}{ab^2}\partial_{t}(-v_2-2\alpha^2c^2)l_y^2\notag\\
&\quad+2\alpha^2\frac{\partial_{t}a}{ab^2}\partial_{t}(-v_3-2\alpha^2c^2)l_z^2\notag\\
&\quad+\alpha^2\bigl(\frac{\partial_{t}a}{ab^2}\partial_{t}(-4o_1+j_1)+2\alpha^2\frac{1}{a^6}\bigr)l_yl_z\notag\\
&\quad+\alpha^2\bigl(\frac{\partial_{t}a}{ab^2}\partial_{t}(-4o_2+j_2)+2\alpha^2\frac{1}{a^6}\bigr)l_zl_x\notag\\
&\quad+\alpha^2\bigl(\frac{\partial_{t}a}{ab^2}\partial_{t}(-4o_3+j_3)+2\alpha^2\frac{1}{a^6}\bigr)l_xl_y\Bigr),\\
8\pi p&=-\frac{1}{a^2\partial_{t}a}\partial_{t}\frac{a(\partial_{t}a)^2}{b^2},
\end{align}
where $c$ is defined as above. One can observe that the function $-4o_n+j_n$ satisfies the same equation as the function $v_n$
\begin{equation}
\partial_{tt}(-4o_n+j_n)-\frac{\partial_{t}\frac{b}{a^3}}{\frac{b}{a^3}}\partial_{t}(-4o_n+j_n)+2\alpha^2\frac{b^2}{a^6}=0,
\end{equation}
so when the scale factor is a power function of time, then the energy density is also a power function of time even though the metric is essentially not.

\end{document}